\documentclass[preprint,showpacs,preprintnumbers,floatfix]{revtex4}
\usepackage{graphicx}
\usepackage{dcolumn}
\usepackage{bm}
\usepackage{longtable}

\begin{document}
\title{Invariance of the Kohn (sloshing) mode in a conserving theory}

\author{M.~Bonitz$^{1}$, K.~Balzer$^{1}$, and R.~van Leeuwen$^2$}
\affiliation{$^{1}$Christian-Albrechts-Universit{\"a}t zu Kiel,
Institut f{\"u}r Theoretische Physik und Astrophysik, Leibnizstr. 15, 24098 Kiel, Germany}
\affiliation{$^2$ Zernike Institute for Advanced Materials, University of Groningen, Nijenborgh 4, 9747 AG Groningen, The Netherlands}
\date{\today}

\begin{abstract}
It is proven that the center of mass (COM or Kohn) oscillation of a many-body system in a harmonic trap coincides with the motion of a single particle as long as conserving approximations are applied to treat the interactions. The two conditions formulated 
by Kadanoff and Baym \cite{kb-book} are shown to be sufficient to preserve the COM mode. The result equally applies to zero and finite temperature, as well as to nonequilibrium situations, and to the linear and nonlinear response regimes.
\end{abstract}
\pacs{05.30.-d, 05.60.Gg, 03.75.Kk}
\maketitle

Interacting many-body systems in confinement potentials are of growing interest in 
many fields, including ions in Penning or Paul traps \cite{itano,Drewsen}, dusty plasmas \cite{arp04,bonitz-etal.06prl}, electrons 
and excitons in quantum wells and dots, e.g. \cite{brey89,afilinov-etal.01prl}
or ultracold atoms and molecules forming Bose condensates, e.g. \cite{pethick02,ohashi04}. In all these systems the problem of collective modes is of high interest both, from the fundamental point of view of the $N$-particle 
spectrum and for the computation of optical and transport properties, e.g. \cite{kohn61,brey89}. The normal mode spectrum has turned 
out to be an important and sensitive test of the quality of theoretical methods in many-body theory, e.g. \cite{ohashi04,reidl01,kwong-etal.00prl} and of their numerical accuracy.

Among all 
collective modes, the center of mass (COM) or dipole mode (Kohn or sloshing mode)
plays a central role.
The dynamics of this mode in an $N$-particle system can be efficiently analyzed using the familiar 
operators \cite{kohn61,brey89}
\begin{eqnarray}
\hat{C}^\pm&=& \sum_{k=1}^N m\Omega {\hat x}_k  \mp i{\hat p}_k,
\label{cpm}
\end{eqnarray}
where ${\hat x}_k$ and ${\hat p}_k$ denote coordinate and momentum of particle $k$ and, to simplify the notation, we 
restrict ourselves to the one-dimensional case (the extension to 3D is straightforward and will be performed below).
Application of $\hat{C}^+$ to the ground state creates an excited state with excess energy $\hbar \Omega$ corresponding 
to an oscillation of the whole system in the confinement potential with the trap frequency $\Omega$. In the case of 
non-interacting particles all oscillate with the same frequency, and no deformation of the density profile occurs. 
Mathematically, this statement follows from the Heisenberg equation of motion for the operators $\hat{C}^\pm$, 
\begin{eqnarray}
\frac{d}{dt}\hat{C}^\pm = \frac{i}{\hbar}[{\hat H}_1,\hat{C}^\pm]  = \pm i \Omega\hat{C}^\pm,
\label{comm}
\end{eqnarray}
showing that, as a function of time, $\hat{C}^\pm$ perform only oscillations without change of the amplitude.
Here ${\hat H}_1 = \sum_k \left( -\frac{\hbar^2}{2m} \partial^2_{x_k} + V(x_k) \right)$ is the single-particle contribution 
to the hamiltonian involving the external harmonic potential $V(x)=m\Omega^2 x^2/2$.

It is well known \cite{kohn61} that 
this behavior is not altered by an arbitrary pair interaction $w(x_i-x_j)$ between the particles leading to 
the interacting hamiltonian ${\hat H} ={\hat H}_1 + V_{12}$, with $V_{12}=\sum_{i<j} w(x_i-x_j)$. One readily verifies that 
$[{\hat V}_{12},\hat{C}^\pm]=0$ which is a direct consequence of translational invariance of the interaction.  Thus, the 
equation of motion (\ref{comm}) remains true also for an interacting $N$-body system if the interactions are treated 
exactly which will be called ``Kohn theorem''. 
Note that this is not sufficient for invariance of the density profile $n({\bf r},t)$ which requires additional 
restrictions on the initial profile $n({\bf r},t=0)$ which will be discussed below.

Obviously, in many situations of practical interest, an approximate treatment of interaction contributions cannot be 
avoided. The simplest model is to approximate $V_{12}$ by an effective mean field (Hartree approximation), 
$V_{12}\rightarrow U_H(x)=\int dx' w(x-x')n(x')$, where $n(x)$ is the density. Apparently, $U_H$ does not possess 
translational invariance and it has been stated \cite{ohashi04} that $[U_{H},\hat{C}^\pm] \ne 0$ leading to a modification of the 
COM mode. This statement is wrong as can be verified by a lengthy but straightforward calculation \cite{rvl_tobe}. In recent years, many 
improved approximations have been discussed in a variety of fields, including the Hartree-Fock approximation, Hartree-Fock-Bogolyubov 
generalized random phase approximation (RPA) \cite{ohashi04} or the time-dependent local density approximation \cite{vignale95}, and rather sophisticated proofs of preservation of the Kohn mode in linear response have been presented, e.g. \cite{reidl01}.

Here, we show that preservation of the Kohn mode can be proven for any approximation which conserves particle number and the total momentum. We further demonstrate that the result holds independently of whether the system was initially 
in the ground state, in equilibrium or nonequilibrium. Finally, the result equally holds for linear response 
and for the case of arbitrary strong excitation.

A very compact proof can 
be obtained using the formalism of nonequilibrium Green's functions, e.g. 
\cite{kb-book,keldysh64,bonitz-book}. To this end we use the second quantization representation of 
the operators $\hat{C}^\pm$ in terms of (fermionic or bosonic) field operators 
\begin{eqnarray}
\hat{c}^\pm(X,T)&=&\left\{\hat{\gamma}^\pm(x_1,x_1')\,\Psi^\dagger(x_1', t_1')\,\Psi(x_1,t_1)\right\}_{1=1'}\;,
\label{cpm2}
\end{eqnarray}
with $\hat{\gamma}^\pm(x_1,x_1')\equiv m\,\Omega\,\frac{x_1+x_1'}{2}\mp\,
\hbar\frac{\partial_{x_1}-\partial_{x_1'}}{2}$. After action of $\hat{\gamma}$ the 
 result is taken at equal arguments $1\equiv (x_1,t_1)=1'\equiv (x_1',t_1')$ which 
equal the macroscopic (center of mass) coordinate and time, $X\equiv (x_1+x_1')/2$, 
$T\equiv (t_1+t_1')/2$. Below, the central quantity of interest, which takes over 
the role of $\hat{C}^\pm$ in Eq. (\ref{comm}), is the ensemble average of (\ref{cpm2}) which is expressed by 
the nonequilibrium Green's function, 
$i\hbar G(1,1')=\langle {\hat T}_c \Psi(1)\Psi^\dagger(1') \rangle$ defined on the 
Schwinger/Keldysh contour $\cal C$, and ${\hat T}_c$ denotes time ordering on $\cal C$ 
\cite{keldysh64}:
\begin{eqnarray}
\label{expectc}
\langle \hat{c}^\pm(X,T)\rangle &=&\left\{\hat{\gamma}^\pm(x_{1},x_{1'})\,i\hbar\,{\bar G}(1,1')\right\}_{1'=1^+}\;,
\end{eqnarray}
where $1^+$ denotes the limit from above for the time argument, $t_1^+=t_1+0$. To simplify the notation in the derivations 
below which apply to fermions and bosons it is helpful to introduce ${\bar G}$ which equals $G$ for bosons and $-G$ for fermions.

To analyze the behavior of the COM mode of the many-body system as a whole, we need to 
compute the time dependence of the space integrated expectation value 
\begin{eqnarray}
\label{expectc_int}
{c}^\pm(T)\equiv \int dX \langle \hat{c}^\pm(X,T)\rangle.
\end{eqnarray}
This time dependence follows directly from the 
equations of motion of the Green's functions and the adjoint equation (first equation 
of the Martin Schwinger (MS) hierarchy \cite{msh})
\begin{eqnarray}
\label{KKBE01}
\left\{i\hbar\,\partial_{t_1}-H_1(1)\right\}{\bar G}(1,1') &=&
i\hbar\int_{\cal C} d2\,W(1-2)\,G_{12}(1,2;1'2^+) \pm \delta_C(1-1'),\\
\label{KKBE02}
\left\{-i\hbar\,\partial_{t_{1'}}-H_1(1')\right\}{\bar G}(1,1') &=&
i\hbar\int_{\cal C} d2\,W(1'-2)\,G_{12}(1,2;1',2^+) \pm \delta_C(1-1')\;,
\end{eqnarray}
where $H_1$ is the single particle hamiltonian defined above and $W(1-2)=w(x_1-x_2)\,\delta_C(t_1-t_2)$, with $\delta_C$ being the delta function defined on ${\cal C}$. 

Note that details of the time contour ${\cal C}$ and of the matrix structure of $G$ are 
not important for the derivations below. We only recall the physical content of 
Eqs. (\ref{KKBE01}, \ref{KKBE02}): while the left hand sides describe the single particle 
dynamics, interaction effects (mean field plus correlations) are contained in the 
right hand sides. Here, the central quantity is the two-particle Green's function 
$G_{12}$ which is a generalization of the standard pair distribution function or 
of the binary correlation operators of the BBGKY hierarchy (the BBGKY hierarchy follows from the MS hierarchy in the equal time limit, e.g. \cite{bonitz-book}). 
Practically all many-body approximations can be formulated in terms of $G_{12}$. For example, the Hartree Fock approximation follows 
by substituting
$G_{12}(1,2;1',2')\rightarrow G(1,1')G(1,1')\pm G(1,2')G(2,1')$.
Finally, the advantage of the formalism of Eqs. (\ref{KKBE01}, \ref{KKBE02}) is that it allows to 
generalize all approximations known from ground state many-body theory, including diagram expansions, to 
arbitrary nonequilibrium situations.

We now proceed to compute the time dependence of ${c}^\pm(T)$, Eq. (\ref{expectc_int}),
and show that one exactly recovers equation (\ref{comm}). In particular, we  
show that the contributions from $G_{12}$, i.e. from the interactions, to the dynamics 
of ${c}^\pm(T)$ vanish exactly if (A): the Green's function $G$ obeys the two mutually adjoint equations (\ref{KKBE01}, \ref{KKBE02}) and (B): for $G_{12}$ a conserving approximation is used which only requires $G_{12}(1,2;1',2')=G_{12}(2,1;2',1')$. These are exactly the two criteria of Baym and Kadanoff known to be sufficient for conservation of density (continuity equation), 
momentum and total energy \cite{baym61,kb-book}.

The time derivative of ${c}^\pm(T)$ is computed directly using the definitions 
(\ref{expectc}, \ref{expectc_int}) and taking the difference of Eqs. (\ref{KKBE01}, \ref{KKBE02})
\begin{eqnarray}
&&\frac{d}{dT}{c}^\pm(T) =
\int dX\,\left\{\hat{\gamma}^\pm(x_{1},x_{1'})\,(\partial_{t_{1}}+\partial_{t_{1'}})\,
i\hbar\,{\bar G}(1,1')\right\}_{1'=1^+}\nonumber\\[0.8pc]
&=&-\int dX\,\left\{\hat{\gamma}^\pm(x_{1},x_{1'})\,\left[\hbar^2(\partial_{x_1}-\partial_{x_1'})\,\frac{\partial_{x_1}+\partial_{x_1'}}{2m}-\left[V(x_1)-V(x_1')\right]\right]{\bar G}(1,1')\right\}_{1'=1^+}\nonumber\\
&& + \,i\hbar\int dX\,dx_2\,\left\{\hat{\gamma}^\pm(x_{1},x_{1'})
\left[w(x_1-x_2)-w(x_{1'}-x_2)\right]\,G_{12}(1,2;1',2^+)\right\}_{1'=1^+}\;,
\label{dcdt}
\end{eqnarray}
where the terms $\delta_C$ cancel and, in the interaction term, the time integration 
has been carried out. We now apply the operator ${\hat \gamma}^\pm$ under the integrals 
and obtain the single particle contribution $dc_1^\pm/dT$ to (\ref{dcdt}):
\begin{eqnarray}
\frac{d}{dT}{c}_{1}^\pm (T)&=&
-\int dX\,\bigg\{\left[m\,\Omega\,X\mp \hbar\frac{\partial_{x_1}-\partial_{x_{1'}}}{2}\right]\,\times
\nonumber\\
&&\hspace{3pc}\times\left.
\left[\hbar^2\frac{\partial_{x_{1}}-\partial_{x_{1'}}}{2m}\,
\partial_{X} - \left[V(x_1)-V(x_1')\right]\right]{\bar G}(1,1')\bigg\}\right|_{1'=1^+}
\nonumber
\\
&=& - \int dX\, \bigg\{\bigg[\pm\frac{V'(x_1)+V'(x_1')}{2}
-\hbar \Omega\frac{\partial_{x_1}-\partial_{x_{1'}}}{2}
\bigg]\hbar {\bar G}(1,1')\bigg\}\bigg|_{1'=1^+},
\end{eqnarray}
where a partial integration over $X$ has been performed taking into account that ${\bar G}$ 
vanishes for $|X|\rightarrow \infty$. For a harmonic confinement potential $V$
follows $d c_{1}^\pm/dT = \pm i \Omega c^\pm$. 

Consider now the interaction contribution to (\ref{dcdt}):
\begin{eqnarray}
\frac{d}{dT}c_{12}^\pm(T)
&=&\,i\hbar\int dX dx_2 \bigg\{\left[m\,\Omega\,X \mp \hbar\frac{\partial_{x_1}-\partial_{x_{1'}}}{2}\right]
\nonumber\\
&\times &\left[w(x_1-x_2)-w(x_{1'}-x_2)\right]\,G_{12}(1,2;1',2^+)\bigg\}_{1'=1^+}.
\nonumber
\end{eqnarray}
Contributions proportional to $\Omega X$ vanish because the potentials cancel for 
$x_1=x_1'$. The remaining contribution involves derivatives of the interaction 
potential, and the integrand can be transformed according to
\begin{eqnarray}
&&\left\{\left[\partial_{x_1}w(x_1-x_2)+\partial_{x_{1'}}w(x_{1'}-x_2)\right]\,G_{12}(1,2;1',2^+)\right\}_{1'=1^+}
\nonumber\\
&=&
\left\{2 w'(X-x_2)\,G_{12}(1,2;1',2^+)\right\}_{1'=1^+}
\;.\nonumber
\end{eqnarray}
Since the force $-w'(x)$ is an odd function of the argument, the expression 
in parantheses vanishes under the integral over the coordinates of both particles if $G_{12}$ is even. But this is exactly condition (B).
Thus, symmetry of $G_{12}$ with respect to the arguments of particle $1$ and $2$ 
is sufficient for vanishing of the time derivative $dc_{12}^\pm/dT$. Summarizing 
the above results for $c_1$ and $c_{12}$, we obtain for the sum (\ref{dcdt}) 
\begin{eqnarray}
\frac{d}{dT} c^\pm = \pm i \Omega c^\pm.
\end{eqnarray}
Thus, we exactly recover the dynamics of Eq.~(\ref{comm}) 
with the only assumption that conditions (A) and (B) are fulfilled. 
Note that our derivation did not involve information on the type of pair interaction 
$w$ and is thus valid for arbitrary pair potentials. Also, no specific 
ensemble had to be specified for the average (in the Green's function), therefore,
the derivation is valid for $N$-body systems originally in the ground state, at 
finite temperature or in nonequilibrium.

As was shown by Baym \cite{baym62} conditions (A) and (B)
are equivalent to existence of a functional $\Phi$ such that $\Sigma(1,1')=\delta\Phi[G]/\delta{G(1',1)}$. 
This means we may conclude that any ``$\Phi-$derivable approximation'' for the selfenergy fulfills the Kohn theorem. 
This result has important implications not only for Green functions theory but also for other classes of approximations, 
such as the ones used in time-dependent density functional theory (TDDFT). It was recently proven \cite{barth05} that any exchange 
correlation functional of TDDFT which is derived from a $\Phi-$derivable approximation satisfies the so-called Zero-Force 
Theorem. Thus, conditions (A) and (B) are suitable to construct TDDFT approximations which preserve the Kohn mode \cite{vignale95}.

A number of generalizations are evident. Our result is trivially generalized to the 3D (2D) case with a confinement 
of the form $\sum_{i=1}^M m\Omega_i^2x_i^2$, where $M=3 (2)$. In this case, there exist three (two) 
independent COM oscillations, and the derivations can be repeated separately for each 
of the three (two) expectation values $c_i(T)$ which now involve $\Omega_i$ instead 
of $\Omega$. Also, while our derivation was performed for a single component system, generalizations 
to multicomponent systems, including Bose condensates, superfluid Fermi gases or mixtures of 
fermions and bosons, should be straightforward.
 Furthermore, the derivations presented in this work are readily generalized to systems in an additional one-particle potential 
$U({\bf r},t)=U_0(t)+U_1(t){\bf r}$, such as an electromagnetic 
dipole potential ($U_0$ and $U_1$ are spatially homogeneous), since this only leads to a 
replacement $V({\bf r}) \rightarrow V({\bf r})+U({\bf r},t)$ in the single particle hamiltonian $H_1$.
In that case, the equations of motion (\ref{comm}) of the operators $C^{\pm}$ are changed to
$\frac{d}{dt}\hat{C}^\pm = \pm i \Omega\hat{C}^\pm \pm i N U_1(t)$. From this it follows that the 
center of mass $R=\frac{1}{N}\sum_k x_k = (C^+ + C^-)/(2mN\Omega)$ obeys the equation 
${\ddot R}+\Omega^2 R=-U_1(t)/m$, which is just the classcial equation of the forced harmonic oscillator. Again, 
this result is not altered by interactions as long as they are treated within approximations obeying 
conditions (A) and (B).

Our result can also be directly extended to 
other types of collective modes. As an example, we mention the cyclotron 
oscillations of charged particles in a homogeneous magnetic field $(0,0,B)$. Then, the 
operators ${\hat C}^{\pm}$ describing the modes have to be replaced by \cite{kohn61}
$P_{\pm}=P_x\pm i P_y$ with ${\bf P}=(p_x,p_y+\frac{eB}{c},p_z)$. Again, it is 
known that the cyclotron rotation with $\omega_c=eB/(mc)$ (which is excited by 
action of $P^+$ on the ground state) is not altered by pair 
interactions if they are treated exactly \cite{kohn61}. The question of invariance 
of the cyclotron rotation under an approximate treatment of the interaction can be 
analyzed exactly as above. Instead of the operators $\partial_{x_1}-\partial_{x_1'}$ 
now there will be linear combinations of different spatial components of the 
momentum operator which again lead to vanishing of the interaction contributions 
$dc_{12}/dT$, for any conserving approximation. 

Finally, let us come back to the question under what conditions the density profile $n({\bf r},t)$ will remain 
unchanged (in the frame oscillating with the center of mass). 
It is easily verified that invariance of $n({\bf r},t)$ in an interacting system is a direct consequence of the 
fulfillment of Eq.~(\ref{comm}), if initially the system was prepared in an eigenstate of ${\hat H}$ or 
in a (correlated) thermodynamic equilibrium state \cite{baym62}. 

In summary, we have shown that the center of mass oscillation (Kohn or sloshing mode)
of interacting particles in a harmonic potential remains unchanged also in an 
approximate treatment of the many-body problem, for any conserving approximation.
This result is of fundamental importance for many fields of physics where trapping 
potentials are being used, including ultracold ions, dusty plasmas, nanostructures or 
quantum gases. 
For theoretical treatments, our result allows to single out the 
scope of physically reasonable approximations, which are just the $\Phi-$derivable 
approximations of Baym. At the same time, it predicts that the shape of 
a density distribution of interacting classical or quantum particles in a harmonic 
trap should remain unchanged in time, if initially only the center of mass oscillation 
is excited and the initial state was either an eigenstate of the hamiltonian or equilibrium. 

\begin{acknowledgments}
We acknowledge stimulating discussions with D. Pfannkuche.
This work is supported by the Deutsche Forschungsgemeinschaft via SFB-TR 24 and by the Innovationsfond Schleswig-Holstein.
\end{acknowledgments}


\begin{thebibliography}{20}


\bibitem{kb-book} L.P.~Kadanoff, and G.~Baym, {\em Quantum Statistical Mechanics}, 
W.A. Benjamin, New York 1962

\bibitem{itano} D.J. Wineland, J.C. Bergquist, W.M. Itano, J.J. Bollinger, and
C.H. Manney, Phys. Rev. Lett. {\bf 59}, 2935 (1987)

\bibitem{Drewsen} M.~Drewsen et al.,
Phys. Rev. Lett. {\bf 81}, 2878 (1998)

\bibitem{arp04}
O.~Arp, D.~Block, A.~Piel, and A.~Melzer, Phys. Rev. Lett. {\bf 93}, 165004 (2004)

\bibitem{bonitz-etal.06prl} M.~Bonitz, D.~Block, O.~Arp, V.~Golubnychiy, H.~Baumgartner, P.~Ludwig, A.~Piel, and A.~Filinov,
Phys. Rev. Lett. {\bf 96}, 075001 (2006)

\bibitem{brey89} L. Brey, N.F. Johnson, and B.I. Halperin, Phys. Rev. {\bf B 40}, 10647 (1989)

\bibitem{afilinov-etal.01prl} A. Filinov, M. Bonitz, and Yu. Lozovik, Phys. Rev. Lett. 
{\bf 86}, 3851 (2001)

\bibitem{ohashi04} Y. Ohashi, Phys. Rev. {\bf A 70}, 063613 (2004)

\bibitem{pethick02} C.J. Pethick, and H. Smith, {\em Bose-Einstein Condensation in Dilute Gases}, Cambridge, New York, 2002

\bibitem{kohn61} W.~Kohn, Phys. Rev. {\bf 123}, 1242 (1961)

\bibitem{reidl01} J.~Reidl, G.~Bene, R. Graham, and P.~Szepfalusy, Phys. Rev. {\bf A 63}, 043605 (2001)

\bibitem{kwong-etal.00prl} N.H.~Kwong, and M.~Bonitz, Phys. Rev. Lett. {\bf 84}, 1768 (2000)

\bibitem{rvl_tobe} R. van Leeuwen, K. Balzer, and M. Bonitz, to be published

\bibitem{vignale95} G.~Vignale, Phys. Rev. Lett. {\bf 74}, 3233 (1995)

\bibitem{keldysh64} L.V. Keldysh, ZhETF {\bf 47}, 1515 (1964) [Sov. Phys. JETP {\bf 20}, 235 (1965)]

\bibitem{msh} P.C. Martin, and J. Schwinger, Phys. Rev. {\bf 115}, 1342 (1959)

\bibitem{baym61} G.~Baym, and L.P.~Kadanoff, Phys. Rev. {\bf 124}, 287 (1961)

\bibitem{baym62} G.~Baym, Phys. Rev. {\bf 127}, 1391 (1962)

\bibitem{bonitz-book} M.~Bonitz, {\em Quantum Kinetic Theory}, B.G. Teubner, Stuttgart/Leipzig 1998

\bibitem{barth05} U. von Barth, N.E. Dahlen, R. van Leeuwen, and G. Stefanucci, Phys. Rev. B {\bf 72}, 235109 (2005)



\end{thebibliography}
\end{document}